\documentclass[aps,pre, twocolumn,showpacs,floatfix]{revtex4}

\usepackage{graphics}
\usepackage{graphicx}
\usepackage{epsfig}
\usepackage{graphics}
\usepackage{dcolumn}
\usepackage{bm}
\begin{document}
\title{Exploring the operation of a tiny heat engine}
\author{Mesfin Asfaw}
\altaffiliation{ Present address: Max Planck Institute of Colloids and
Interfaces, 14424 Potsdam, Germany}
\email{asfaw@mpikg.mpg.de}
\affiliation{Department of physics, Addis Ababa University\\
 P.O.Box 1176, Addis Ababa, Ethiopia}

\author{Mulugeta Bekele}%
\email{mbekele@phys.aau.edu.et}
\affiliation{Department of physics, Addis Ababa University\\
 P.O.Box 1176, Addis Ababa, Ethiopia}

\date{\today}

\begin{abstract}
We model a tiny heat engine as a Brownian particle that moves in a viscous medium in a sawtooth potential (with or without load) assisted by $\it {alternately}$ placed hot and cold heat baths along its path. We find closed form expression for the steady state current as a function of the model parameters. This enables us to deal with the energetics of the model and evaluate either its efficiency or its coefficient of performance depending upon whether the model functions either as a heat engine or as a refrigerator, respectively. We also study the way current changes with changes in parameters of interest. When we plot the phase diagrams showing the way the model operates, we not only find regions where it functions as a heat engine and as a refrigerator but we also identify a region where the model functions as neither of the two.
\end{abstract}
         
 \maketitle

 \section  { \bf Introduction}

Understanding the physical properties of devices of micron- and nano-meters sizes is of interest these days since the trend is in miniaturizing them. Such devices may need to be kept under certain environmental condition such as a particular temperature in order for them to optimally operate.  We envision heat pumps (or refrigerators) of similar sizes that provide such favorable condition. On the other hand, heat engines of similar sizes may provide energy to these devices in order for them to perform different tasks. This paper deals with a model of such tiny heat engine or, conversely, heat pump and explores the details. 

The idea of microscopic (or Brownian) heat engine working due to non-uniform temperature first came up with the works of B{\"u}ttiker \cite{but}, van kampen \cite{van}, and Landuaer \cite{lan1} while they were involved in exposing the significance of the now influential papers of Landauer on blowtorch effect \cite{lan2,lan3}. Later works took models of Brownian heat engine and dealt with the energetics only at the quasi-static limit \cite{miki3,Aus1}. Energetics considerations for such engine operating in a {\t finite time} were first addressed by Der\`enyi et al.  \cite{Ast2} and recently by us \cite{mesfin1,mesfin2}. These two works of us also revealed that the quasi-static limits of either the heat engine or the heat pump behave exactly like that of a Carnot's heat engine or heat pump.

In our first paper \cite{mesfin1}, we considered a Brownian particle moving through a highly viscous medium in a periodic sawtooth potential (with or without load) assisted by the thermal kick it gets from {\it alternately} placed hot and cold heat reservoirs along its path. The heat reservoirs were placed in such a way that the whole left side of each sawtooth from its barrier top is coupled to the hot reservoir while the whole right side is coupled to the cold reservoir.

In the present work, instead of coupling the whole left side of each sawtooth to the hot reservoir we let it be hot within a narrower but varying width located at a position that can also vary as a parameter. This will introduce two additional parameters to the model and, thereby, address a more general problem. By evaluating the way energy exchange takes place between the particle and the medium, we explore the conditions under which the model works either as a heat engine or as a heat pump. One new result is the existence of a third region in the parameter space where the model works neither as a heat engine nor as a heat pump with rich phase diagrams. 

The paper is organized as follows:  In section II we will consider our model in the absence of external load. In this case we will show that 
the model works only as a heat engine and, at quasistatic limit, the efficiency goes to that of Carnot efficiency. In section III  we will consider our model in the presence of external load and explore how current, efficiency and coefficient of performance of the model vary with change in model parameters of interest.   We will identify the  different operating regions (as a heat engine,  as a refrigerator or neither of the two)  of the model in phase diagrams. Section IV deals with summary and conclusion.

    \section{The model with no external load}

The model consists of a Brownian particle which moves in a viscous medium in a   periodic sawtooth potential where the background temperature
is also periodic but piecewise constant  as shown in Fig. 1. Here
U(x) denotes the potential of the sawtooth with a barrier height of $U_{0}$. One
of the maxima of the potential is located at x=0 and the two minima on the left and right sides of this maximum potential  are located at $x=-L_{1}$ and $x=L_{2}$, respectively.
 Two dimensionless parameters, $\alpha $ and $\delta $, specify the position and width of each  hot locality such that  $\alpha $
 describes the position of the left side of the hot locality from the nearest
 barrier top while  $\delta  $ describes the width of the hot locality, both in units of $L_{1}$. Here one should note that $\alpha$ and $\delta$ are limited to take values between zero and one: $0<\alpha<1$ and $0<\delta <1$. On the other hand, in our previous work \cite{mesfin1} both $\alpha$ and $\delta$ are taken to have a fixed value of one. The quantities $T_{h}$  and $T_{c}$ in the figure denote  the temperature  values of the hot and cold  localities, respectively.
 A third parameter, $\tau$, relates the temperature of the hot locality with the cold locality such that $T_{h}=T_{c}(1+\tau ) $. 
\begin{figure} 
\epsfig{file=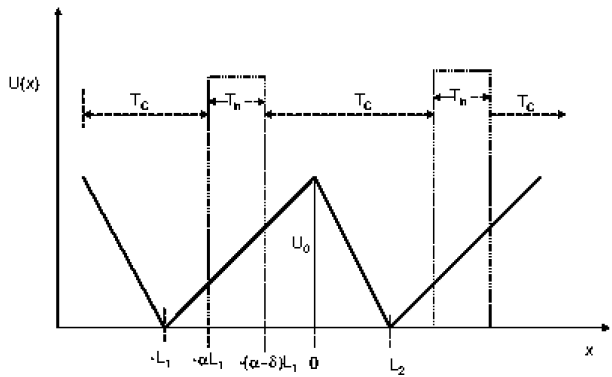,width=8cm}
\caption{The plot of periodic sawtooth potential without load. The parameter $\alpha$ describes the left side of hot locality  from the nearest barrier top while $\delta$ describes the width of hot locality both in the units of $L_{1}$.
When  $\alpha=1$ and $\delta=0$ the sawtooth potential are coupled to the cold reservoir while when  $\alpha=1$ and $\delta=1$ the left part of the  sawtooth potential are coupled with the hot reservoir. }
\end{figure}
The potential, $U(x)$, for our model is a series of identical sawtooth potentials.
  One sawtooth potential  around x=0 is described by
  \begin{equation} U(x)=\cases{
   U_{0}[{x\over L_{1}}+1],&if $-L_{1}< x \le 0$;\cr
   U_{0}[{-x\over L_{2}}+1],&if $0 < x \le L_{2}$;\cr
   }\end{equation} 
    and this potential profile repeats itself such that $U(x+L)=U(x)$ where $L=L_{1}+L_{2}$.
   Note that $U(x)$ is determined by three parameters: $U_{0}$, $L_{1}$ and $L_{2}$.
   On the other hand, the temperature profile  is given by 
   \begin{equation} T(x)=T_{c}+\tau T_{c} (\Theta (x-\alpha L_{1})- \Theta (x-(\alpha+\delta) L_{1}), \end{equation} 
   for $-L_{1}<x \le L_{2}$ and  $\Theta$ is the Heavyside function. This temperature profile repeats itself with the same period as that of the potential: $T(x+L)=T(x)$.

The presence of hot and cold regions coupled to the different parts of the sawtooth potential determines the Brownian particle to be driven unidirectionally and attain a steady state current, $J$. A general expression for the steady state current of the Brownian particle in any periodic potential (with or without load) is derived in Appendix A.
Using the particular potential and temperature profiles of Eqs. (1) and (2), we have evaluated the expressions for $F$, $G_1$, $G_2$ and $H$ in Appendix A, Eqs. (A10, A11,A12, and A13), and
found  an exact  closed form expression for the steady state current $J$ 

     \begin{equation}  J={-F\over G_{1}G_{2} +H F}. \end{equation} 

The expressions for $F$, $G_{1}$, $G_{2},$ and  $H$ are given in Appendix B.
The  drift velocity, $v$, is  then given by  \begin{equation} v=J(L_{1}+L_{2}).\end{equation} 

As an approximation, we neglect the energy transfer via kinetic energy due to the particle recrossing of the boundary between the regions \cite{Aus1,Ast2} and find the energetics of the particle as it exchanges energy with the heat reservoirs.
In one cycle the Brownian particle is in contact with the hot region of width  $\delta L_{1}$ while the particle is in contact with the cold region of width $L_{2}+L_{1}(1-\delta)$. When the particle moves through the hot region, it takes an energy $Q_{h}$ which enables it not only to climb up the potential of magnitude $\delta U_{0}$ but also  acquire energy $\delta v\gamma L_{1}$ to overcome the viscous drag force within the hot interval. 
 Hence during  one  cycle, the amount of heat flow out of the hot reservoir, $Q_{h}$, is given by
 \begin{equation}  Q_{h}=\delta U_{0}+\gamma v L_{1}\delta.\end{equation} 
The net heat  flow per cycle from the Brownian particle to the cold reservoir, $Q_{c}$, is by a similar argument given by 
 \begin{equation} Q_{c}=\delta U_{0}-\gamma v (L_{2}+L_{1}(1-\delta)).\end{equation}
The difference between $Q_{h}$ and  $Q_{c}$  is the useful work:
\begin{equation}
W=\gamma v(L_{1}+L_{2}).
\end{equation}
This is exactly the amount of work required to transport the Brownian particle moving with a drift velocity $v$ through one cycle in the viscous medium. 
It is important to emphasize at this point that in the absence of external load the model works only as heat engine.
Following the definition of generalized efficiency introduced in \cite{Ast2}, the efficiency of the engine, $\eta$, is given by
 \begin{equation} \eta={Q_{h}-Q_{c}\over Q_{h}}={\gamma v (L_{1}+L_{2})\over \delta (U_{0}+\gamma v L_{1})}.\end{equation} 
One can notice that the magnitude of the current approaches  to zero when  $U_{0}$ goes to zero. This corresponds to the quasistatic limit of the engine; i.e. $J \to 0$. 
Evaluating the efficiency at quasistatic limit, we found that

\begin{equation}  \displaystyle \lim _{{U_{0}\to 0}}{\eta }={(T_{h}-T_{c})\over   T_{h}}, \end{equation} 
 which is exactly equal to the efficiency of a Carnot engine.

\begin{figure} 
\epsfig{file=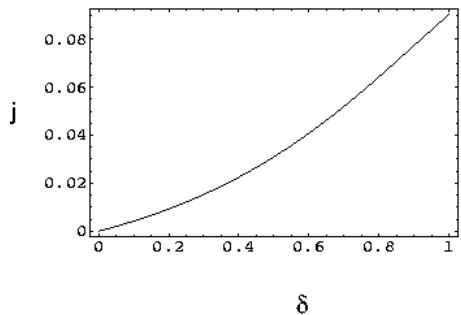,width=6cm}
\caption{ Plot of $j$ versus $\delta$  for fixed  $\tau$=1, $\ell$=$3$, $u=4$ and $\alpha=1$.}
\end{figure}

\begin{figure} 
\epsfig{file=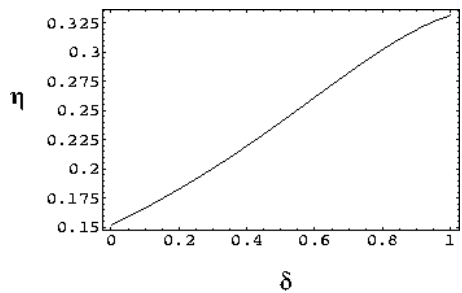,width=6cm}
\caption{ Plot of $\eta$ versus $\delta$ for fixed  $\tau$=1, $\ell$=$3$, $u=4$ and $\alpha=1$.}
\end{figure}

In order for us to explore how current and efficiency vary with change in one of the newly introduced parameters $\alpha$ and $\delta$, we first introduce three scaled parameters: scaled length
 $\ell=L_{2}/ L_{1} $, scaled barrier height $ u=U_{0}/ k_BT_{c}$ and scaled current $j=J/J_{0}$ where $J_{0}=k_BT_{c}/\gamma L_{1}^2$; $k_B$ being Boltzmann's constant.

 For fixed values of  $\alpha=1$, $u=4$ and  $\tau=1$, the current $j$ is plotted as a function of $\delta$ as shown in Fig. 2. When $\delta=0$,
 the whole sawtooth potential is coupled with cold reservoir. Hence there is no net current as can be seen in the figure.  Increasing the width of the hot locality leads to an increase in the current.
 We also plot the engine efficiency, $\eta$, as a function of $\delta$ as shown in Fig. 3. The figure shows that $\eta$  increases with increase in $\delta$. 
\begin{figure} 
\epsfig{file=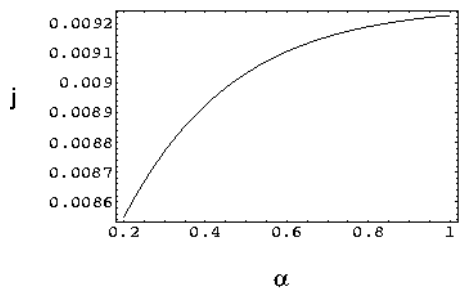,width=6cm}
\caption{ Plot of $j$ versus $\alpha$  for fixed  $\tau$=1, $\ell$=$3$, $u=4$ and $\delta=0.2$.}
\end{figure}

One can fix the width of the hot locality, $\delta$, and investigate the influence of shifting the position of the hot locality along one side of the sawtooth potential. Figure 4 is a plot of $j$ versus $\alpha$ for a given values of $\delta=0.2$, $u=4$ and  $\tau=1$.
The figure shows that the current  $j$ increases as $\alpha$ increases which implies that positioning the hot locality near the potential minimum produces the highest possible value of current than putting it anywhere else. 
Likewise, plot of $\eta$ versus $\alpha$ in Fig. 5 shows similar feature as in Fig. 4 in the sense that one gets highest possible efficiency if the hot locality is put near the potential minimum than anywhere else. 
\begin{figure} 
\epsfig{file=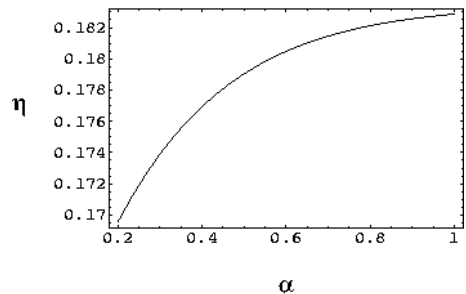,width=6cm}
\caption{ Plot of $\eta$ versus $\alpha$ for fixed  $\tau$=1, $\ell$=$3$, $u=4$ and $\delta=0.2$.}
\end{figure}
 
In the next section we will consider our model to also have external load.

   \section{The model with constant external load}

   Consider the model in the presence of a constant external load $f$ as shown in  Fig. 6.
   The potential profile will then change from  $U(x)$ described by Eq. (1) to $U(x)+fx$.
  \begin{figure} 
\epsfig{file=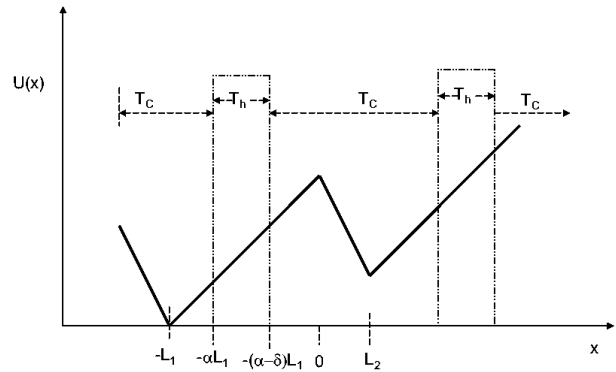,width=8cm}
\caption{Plot of periodic  sawtooth potential with load where periodic hot and cold
 locality alternate along the potential. }
\end{figure}
Similar to  the previous section, the closed form expression for steady state  current $J^{L}$ in the presence of external load
is derived and given by
     \begin{equation}  
     J^{L}={-F^{L}\over G_{1}^{L}G_{2}^{L} +H^{L}F^{L}}.
     \end{equation} 
The expressions for $F^{L}$, $G_{1}^{L}$, $G_{2}^{L},$ and  $H^{L}$ are given in Appendix C.
The  drift velocity, $v^{L}$, is  then given by  \begin{equation} v^{L}=J^{L}(L_{1}+L_{2}).\end{equation} 

\begin{figure} 
\epsfig{file=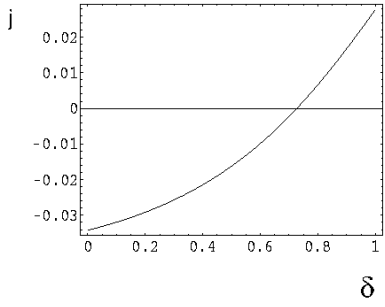,width=5cm}
\caption{ Plots of $j$ versus $\delta$ for fixed  $\tau$=1, $\ell=3$ ,$u=4$, $\lambda=0.4$ and $\alpha=1$.}
\end{figure}
Before we begin exploring various properties of the model, we introduce a scaled quantity associated with the load: $\lambda=fL_{1}/T_{c}$. 
We will then have a total of six parameters that can be controlled independently: $u$, $l$, $\tau$, $\delta$, $\alpha$ and $\lambda$.
Fig. 7 is a plot of (scaled) current $j$ versus $\delta$ for fixed values of all the other parameters. The figure shows that the current monotonously increases with $\delta$ being negative for 
small values of $\delta$ (hot locality width) and then becoming positive beyond a certain value of $\delta$. This implies that in the interval where the current is negative external work is being done {\it on} the engine 
while in the interval where the current is positive the engine {\it does} work not only against the viscous medium but also against the load. 
In order to find whether the engine works either as a heat engine, or as a refrigerator or other wise, we need to find energy exchange between the particle and the heat reservoirs. 
When the model behaves as a heat engine, the current  takes postive value (see Fig. 7)
and in one cycle the particle takes heat energy $Q_{h}^L$ from the hot reservoir of amount
\begin{equation} Q_{h}^{L}=\delta(U_{0}+fL_{1}+\gamma v^L L_{1}),\end{equation} 
and gives off an amount of heat 
  \begin{equation} 
  Q_{c}^{L}=\delta U_{0}-[((L_{1}(1-\delta)+L_{2}))(f+\gamma v^L)]
  \end{equation}
  to the cold reservoir.
  The net work, $W$, done by the heat engine in one cycle is given by the difference between $Q_{h}^L$ and $Q_{c}^L$:
\begin{equation}
W=(\gamma v^L+f)(L_{1}+L_{2}).
\end{equation}
The generalized efficiency of the heat engine will then be given by 
 \begin{equation}
\eta={W\over Q_h^L}={(\gamma v^L+f)(L_{1}+L_{2})\over \delta(U_{0}+fL_{1}+\gamma v^LL_{1})}.
\end{equation}
On the other hand, when the model works as a refrigerator heat flows out of the cold reservoir and enters the hot reservoir driven by the external work of amount $W^L = f(L_1 + L_2)$ done on the refrigerator per cycle. In the region where the model works as a  refrigrator, the coefficient of performance (COP) of the refrigerator, $P_{ref}$ , is given by
\begin{equation}
P_{ref} = { Q_c^L\over W^L}={\delta U_{0}-[((L_{1}(1-\delta)+L_{2}))(f+\gamma v^L)]\over f(L_{1}+L_{2})}.
\end{equation}

The quasi-static limit of the engine corresponds to the case where the current approaches to zero  either from the heat engine
side or from the refrigerator side.  We have found that this limit is satisfied when   
\begin{equation}  
f={\delta\tau U_{0}\over ((1+l)(1+\tau)-\delta\tau)L_{1}}.
\end{equation} 
This is the boundary demarcating the domain of the operation of the engine as a refrigerator from that as a heat engine. 
Evaluating the expression for both $\eta$ and $P_{ref}$ as we approach this boundary, we analytically found that they are exactly equal to that of Carnot efficiency and Carnot COP, respectively: 
\begin{equation}
\displaystyle \lim _{J^+\to 0}{\eta }={(T_{h}-T_{c})\over T_{h}},
\end{equation} 
and 
\begin{equation}
 \displaystyle \lim _{J^-\to 0}{P_{ref} }={T_{c}\over (T_{h}-T_{c})}.
\end{equation}
\begin{figure} 
\epsfig{file=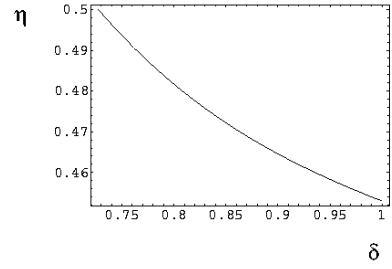,width=5cm}
\caption{Plots of $\eta$ versus $\delta$ for fixed  $\tau$=1, $\ell=3$ ,$u=4$, $\lambda=0.4$ and $\alpha=1$.}
\end{figure}
\begin{figure} 
\epsfig{file=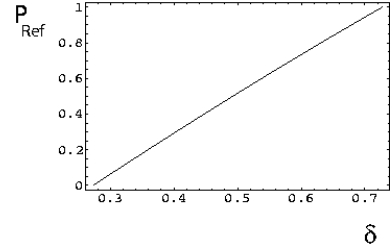,width=5cm}
\caption{Plots of $P_{ref}$ versus $\delta$ for fixed  $\tau$=1, $\ell=3$ ,$u=4$, $\lambda=0.4$ and $\alpha=1$.}
\end{figure}
We now explore how the $\eta$ and $P_{ref}$ behave as a function of $\delta$ fixing all other parameters.
 Within the region where the model works as a heat engine, we find the efficiency to monotonously decrease from its maximum value (Carnot efficiency) as the width, $\delta$, of the hot locality increases ( see  Fig. 8). 
 On the other hand,  within the region where the model works as a refrigerator the COP linearly increases with increase in  $\delta$ until it attains its maximum possible value (Carnot refrigerator) as shown in Fig. 9.

\begin{figure} 
\epsfig{file=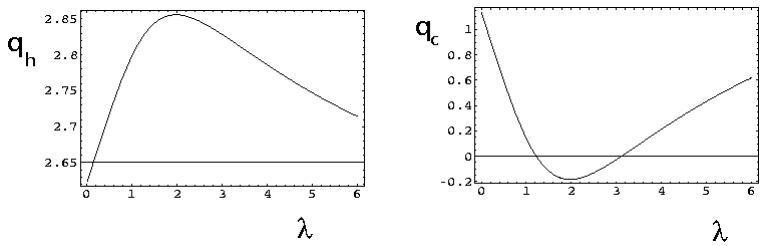,width=10cm}
\caption{(a) Plot $q_{h}$ versus  $\lambda$ for fixed 
  $\tau=4$, $\ell=3$, $u=4$,  $\alpha=1$ and $\delta=0.6$. (b) Plot $q_{c}$ versus  $\lambda$ for fixed 
  $\tau=4$, $\ell=3$, $u=4$, $\alpha=1$ and $\delta=0.6$.}
\end{figure}
Before we plot the phase diagrams showing regions where the model works as a heat engine, as a refrigerator and as neither of the two we think that it is illustrative to know how $Q_h$ and $Q_c$ behave with scaled load, $\lambda$. We introduce two  scaled quantities $q_h=Q_h/k_BT_c$ and $q_c=Q_c/k_BT_c$ corresponding to $Q_h$ and $Q_c$, respectively. Figs. 10a and 10b show plots of $q_h$ versus $\lambda$ and $q_c$ versus $\lambda$, respectively, after fixing the other parameters. 
Note that Fig. 10a shows that $q_{h}$ is always positive while Figure 10b shows that  $q_{c}$ takes a negative value within some interval values of $\lambda$. The quantity $q_{h}-q_{c}$ is always positive. In the region  that $q_{c}$ is negative, the model works neither as a heat engine nor as a refrigerator since under this situation external work is supplying energy to {\it both} the hot and cold reservoirs. 
We  have found that when
 \begin{equation} 0<\lambda <{\delta\tau u\over ((1+l)(1+\tau)-\delta\tau)}\end{equation} 
the model works as a heat engine while the model works as a refrigerator when 
\begin{equation} {\delta\tau u\over ((1+l)(1+\tau)-\delta\tau)}<\lambda<\left({\delta u \over (1-\delta)+l}-j(1+l)\right).\end{equation}
The model works neither as a heat engine nor as a refrigerator if
\begin{equation} 
\lambda>\left({\delta u \over (1-\delta)+l}-j(1+l)\right).
\end{equation}

\begin{figure} 
\epsfig{file=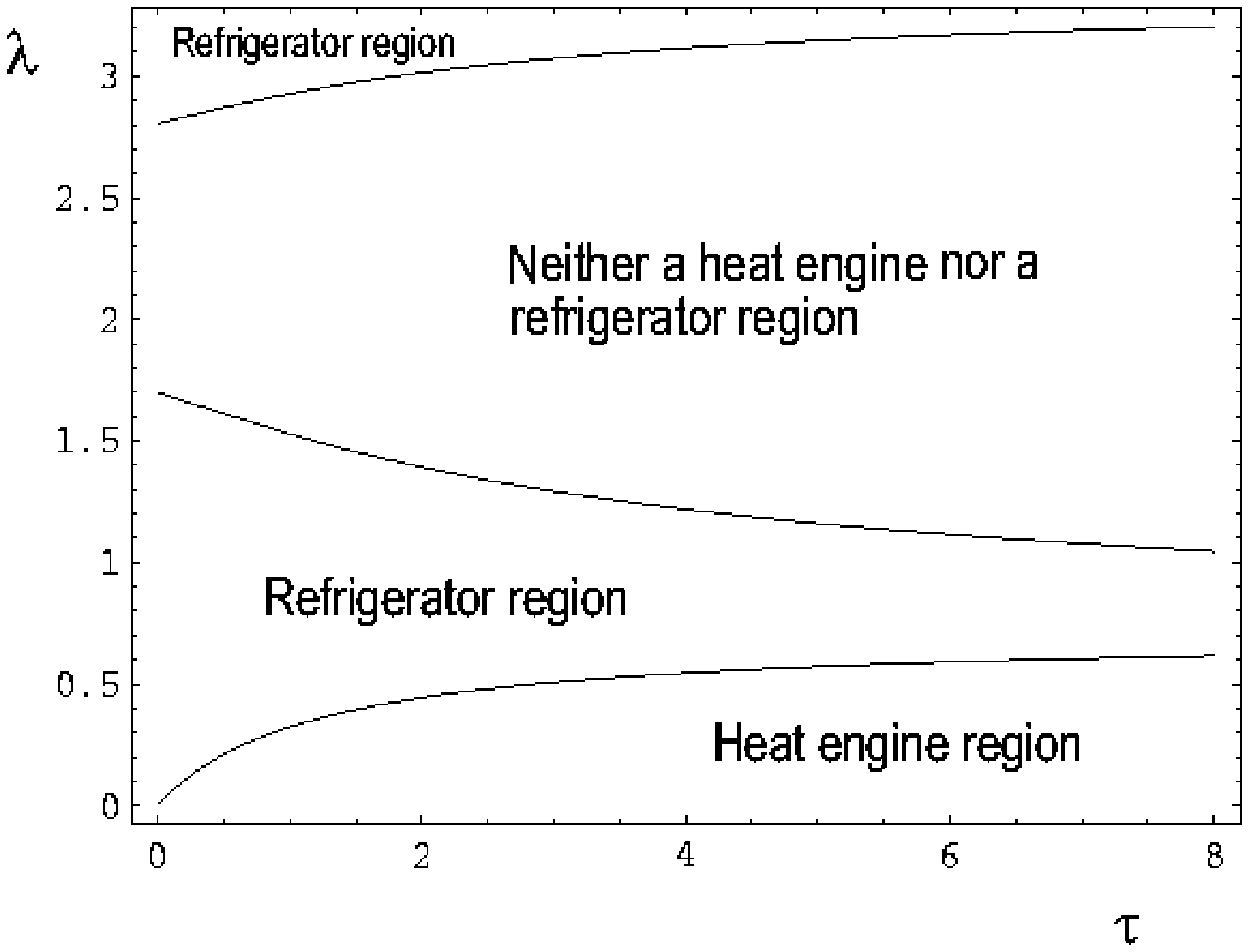,width=6cm}
\caption{ The phase diagram in the $\lambda-\tau$ parameter space  for fixed 
  $\delta=.6$, $\ell=3$, $u=4$ and $\alpha=1$.}
\end{figure}
\begin{figure} 
\epsfig{file=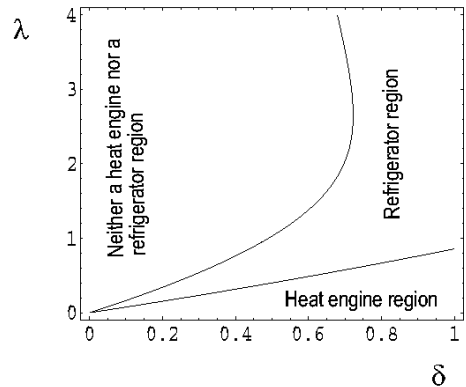,width=6cm}
\caption{ The phase diagram in the $\lambda-\delta$ parameter space  for fixed 
  $\tau$=1, $\ell=3$, $u=4$ and $\alpha=1$.}
\end{figure}
We plot the three operation regions of the model in Fig. 11. The figure demonstrates that in the parameter space of $\lambda - \tau$, the model works as a heat engine, as a refrigerator or as neither of the two for fixed $\delta=.6$, $\ell=3$, $u=4$ and $\alpha=1$. One can see in the phase diagram  that for $\tau=4$, the model works neither as a heat engine nor as a refrigerator when $1.21875<\lambda< 3.156$. In fact in this region, $q_{c}$ is negative as can be seen in Fig.10b.
 Figure 12 shows the three regions in $\lambda - \delta$ parameters space in which the model operates as a heat engine, as a refrigerator and as neither of the two for fixed $l=3$, $\alpha=1$, $\tau=1$and  $u=4$.  

 \section{summary and conclusion}
We considered a simple model of a Brownian heat engine and explored how current, efficiency and performance of a refrigerator behave as model parameters of interest vary. In the absence of external load, the model works as a heat engine  while in the presence of external load the model works  as a heat engine, as a refrigerator  or neither of the two depending on the values in the parameter space. At a quasistatic limit the efficiency as well as the COP of the engine goes to that of Carnot efficiency and Carnot COP. In this paper we reported an extended version  of the  work \cite{mesfin1}. The result we obtained shows that unlike the previous work, there are regions that the model may work neither as a heat engine nor as a refrigerator determined by the width of hot locality.

 In this work, we found a closed expression for  the steady state current. This enabled us to explore the energetics of the Brownian heat engine not only at the quasistatic limits but also while operating at any finite time.
 This is a clear exposition of the power of having analytic expression for the concerned physical quantities.
\appendix
\section {Derivation of the steady state current}
Consider motion of a Brownian particle in any periodic potential, $U(x)$,
and temperature, T(x), profiles where both have the same periodic length of
$L_{1}+L_{2}=L$ as shown in Fig. 13.
\begin{figure}
\epsfig{file=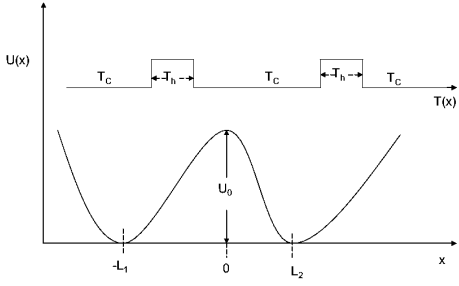,width=6cm}
\caption{Periodic potential with non-homogenous temperature profile. }
\end{figure}
In the high friction limit, the dynamics of the Brownian particle is governed by the Smoluchowski equation \cite{san}:
       \begin{equation} 
       {\partial P(x,t)\over \partial t}={\partial\over  \partial x} ({1\over \gamma }[U'(x)P(x,t)+{\partial \over \partial x}
       (T(x)P(x,t)) ]) 
   \end{equation}
       where P(x,t) is the probability density of finding the particle at position x
        at time t, $U'(x)={d\over dx}U$  and $\gamma $   is the coefficient of friction  of the particle.
         Boltzman's constant, $k_{B}$, and the mass of the particle are taken
         to be unity.
 The steady state solution, $P^{s}(x)$, of the Smoluchowski equation implies a constant current, J, given by
          \begin{equation}                  -{1\over \gamma}[U'(x)P^{s}(x)+{d \over dx}(T(x)
         P^{s}(x))]=J. \end{equation} 
 Periodic boundary condition implies
  \begin{equation} 
  P^{s}(x+L)=P^{s}(x).\end{equation} 
     Equation A3 is valid for any periodic potential  $even$
 $ in$ $the$ $presence$ $of$ $additional$ $constant$  $external$ $load$. 
 The stationary probability distribution can be then  given by
          \begin{equation} 
          P^{s}(x)= {B\over T(x)}exp(-\phi (x))\end{equation} 
where B is the normalization constant and $ \phi (x)=\int _{-L_{1} }^{x} {U'(x)dx'\over T(x')}$
with $-L_{1}<x\le L_{2}$. After multiplying both sides of Eq. (A4) by $ exp(\phi (x))$, we can write it as 
\begin{equation} {d\over dx}[e^{\phi (x)}T(x)P^{s}(x)]={-\gamma J}e^{\phi (x)}.\end{equation} 
Integrating Eq. (A5) from $ -L_{1}$ to $ x $ and, after some algebra, one gets
\begin{equation} P^{s}(x)={e^{-\phi (x)}\over T(x)}{\brack T(-L_{1})P^{s}(-L_{1})
 -  \gamma J \int _{-L_{1}}^{x}  e^{\phi (x)}dx} .\end{equation} 
Taking x=$L_{2}$ in   Eq. (A6), we get
\begin{equation} (e^{\phi (L_{2})}-1)P^{s}(L_{2})T(L_{2})=-\gamma J \int _{-L_{1}}^{L_{2}}
{e^{\phi (x)}}dx.\end{equation} 
 Applying normalizing condition, $ \int _{-L_{1}}^{L_{2}}P^{s}(x)dx=1$, on Eq. (A6) gives us the relation
\begin{eqnarray} 1& = &P^{s}(L_{2})T(L_{2})\int _{-L_{1}}^{L_{2}}{e^{-\phi (x)}\over T(x)}dx- \nonumber \\ & & \gamma J \int _{-L_{1}}^{L_{2}}
{e^{-\phi (x)}\over T(x)} \int _{-L_{1}}^{x}
{e^{\phi (x')}}dx'.\end{eqnarray} 
 Using Eqs. (A7) and (A8) it is simple to get the steady state current 
  \begin{equation} J={-F\over G_{1}G_{2} + HF},\end{equation} 
  where\begin{eqnarray} 
  F&=&e^{\phi (L_{2})}-1,\\
  G_{1}&=&\int _{-L_{1}}^{L_{2}}{e^{-\phi (x) }\over T(x)}dx,\\
  G_{2}&=& \int _{-L_{1}}^{L_{2}}{e^{\phi (x') }\gamma (x')}d'x,\\
   H&=&\int _{-L_{1}}^{L_{2}}{e^{-\phi (x) }\over T(x)}dx
  \int _{-L_{1}}^{x}{e^{\phi (x') }\gamma (x')}d'x.\end{eqnarray}

\begin{widetext}
\section{}
 In this Appendix, we will give the expressions for $F$, $G_{1}$, $G_{2}$, and $H$ which define the value of the steady state current, $J$, for load free case \begin{eqnarray} 
   F& = &e^{\delta U_{0}[{1\over T_{h}}-{1\over T_{c}}]}-1 \\
 G_{1}& = &{L_{1}\over U_{0}}[1-e^{{(\alpha -1)U_{0}\over T_{c}}}]+{L_{1}\over U_{0}}
e^{{(\alpha -1)U_{0}\over T_{c}}}[1-e^{{-\delta U_{0}\over T_{h}}}]+\nonumber \\ & &
{L_{1}\over U_{0}}e^{{(\alpha -1)U_{0}\over T_{c}}-{U_{0}\delta \over T_{h}}}
[1-e^{{(\delta-\alpha) U_{0}\over T_{c}}}]
+{L_{2}\over U_{0}}e^{{(\alpha -1)U_{0}\over T_{c}}-{U_{0}\delta \over T_{h}}-
{(\alpha -\delta) U_{0}\over T_{c}}}[e^{{U_{0}\over T_{c}}}-1] \\ 
G_{2}& =  & {\gamma T_{c}L_{1}\over U_{0}}[e^{{(1-\alpha )U_{0}\over T_{c}}}-1]+ {\gamma T_{c}L_{1}\over U_{0}}e^{{(1-\alpha )U_{0}\over T_{c}}+{U_{0}\delta \over T_{h}}}
[e^{{(\alpha-\delta) U_{0}\over T_{c}}}-1]
+{\gamma T_{h}L_{1}\over U_{0}}e^{{(1-\alpha )U_{0}\over T_{c}}}[e^{{\delta U_{0}\over
 T_{h}}}-1]+{\gamma T_{c}L_{2}\over U_{0}}\nonumber \\ 
& & e^{{(1-\alpha )U_{0}\over T_{c}}+{U_{0}\delta \over T_{h}}+ {(\alpha-\delta) U_{0}\over T_{c}}}
 [1-e^{{-U_{0}\over T_{c}}}]\\
 H& = &t_{1}+t_{2}+t_{3}+t_{4}+t_{5}+t_{6}+t_{7}+t_{8}+t_{9}+t_{10}\\
 t_{1}& = &{L_{1}\gamma \over U_{0}}[(1-\alpha )L_{1}+{T_{c}L_{1}\over U_{0}}[
 e^{{U_{0}(\alpha -1)\over T_{c}}}-1]],\\
  t_{2}& = &{L_{1}^{2}T_{c}\gamma \over U_{0}^{2}}e^{{U_{0}(\alpha -1)\over T_{c}}}
      [e^{{(1-\alpha )U_{0}\over T_{c}}}-1][1-e^{{-U_{0}\delta \over T_{h}}}],\\
 t_{3}& = &{L_{1}\gamma \over U_{0}}[\delta L_{1}+{T_{h}L_{1}\over U_{0}}[e^{{-U_{0}\delta \over T_{h}}}
-1]],\\ 
 t_{4}& = &{L_{1}^{2}T_{c}\gamma \over U_{0}^{2}}[1-e^{{U_{0}(1-\alpha )\over T_{c}}}]
 e^{{U_{0}(\alpha -1)\over T_{c}}-{\delta U_{0}\over T_{h}}}
 [e^{{-U_{0}(\alpha -\delta )\over T_{c}}}-1],\\ 
 t_{5}& = &{L_{1}^{2}T_{h}\gamma \over U_{0}^{2}}[e^{{-U_{0}(\alpha -\delta )\over T_{c}}}-1]
[1-e^{{U_{0}\delta \over T_{h}}}]e^{{-U_{0}\delta \over T_{h}}},\\
t_{6}& = &{L_{1}\gamma \over U_{0}}[(\alpha -\delta )L_{1}+{T_{c}L_{1}\over U_{0}}
[e^{{-U_{0}(\alpha -\delta )\over T_{c}}}-1]],\\ 
 t_{7}& = &{L_{1}L_{2}T_{c}\gamma \over U_{0}^{2}}[e^{{U_{0}(1-\alpha )\over T_{c}}}-1]
e^{{-U_{0}\over T_{c}}+{U_{0}\delta \over T_{c}}-{U_{0}\delta \over T_{h}}}
 [e^{{U_{0}\over T_{c}}}-1],\\
  t_{8}& = &{L_{1}L_{2}T_{h}\gamma \over U_{0}^{2}}[e^{{U_{0}\delta \over T_{h}}}-1]
   [e^{{U_{0}\over T_{c}}}-1]e^{{-U_{0}(\alpha -\delta )\over T_{c}}-{U_{0}\delta \over T_{h}}
   },\\ 
  t_{9}& = &{L_{1}L_{2}T_{c}\gamma \over U_{0}^{2}}e^{{-U_{0}(\alpha -\delta )\over T_{c}}}
  [e^{{U_{0}(\alpha -\delta )\over T_{c}}}-1][e^{{U_{0}\over T_{c}}}-1],\\
  t_{10} & = &{-L_{2}\gamma \over U_{0}}[L_{2}-{L_{2}T_{c}\over U_{0}}[ e^{{U_{0}\over T_{c}}}-1]].\end{eqnarray} 
\section{}
In this Appendix we will give the expressions for $F^{L}$, $G_{1}^{L}$, $G_{2}^{L}$ and   $H^{L}$
  which define the value of the steady state current, $J^{L}$, for nonzero external load case. 
\begin{eqnarray} 
 F^{L}& = &e^{{U_{0}+fL_{1}\over T_{c}}+{-U_{0}+fL_{2}\over T_{c}}+
  {\delta (U_{0}+fL_{1})\over T_{h}}-{\delta (U_{0}+fL_{1})\over T_{c}}}-1,\\ 
G_{1}^{L}& = &{L_{1}\over (U_{0}+fL_{1})}[1-e^{{(\alpha -1)(U_{0}+fL_{1})\over T_{c}}}],
+{L_{1}\over (U_{0}+fL_{1})}
e^{{(\alpha -1)(U_{0}+fL_{1})\over T_{c}}}[1-e^{{-\delta (U_{0}+fL_{1})\over T_{h}}}]
+{L_{1}\over(U_{0}+fL_{1})}e^{{(\alpha -1)(U_{0}+fL_{1})\over T_{c}}-{(U_{0}+fL_{1})\delta \over T_{h}}}\nonumber \\ & &
[1-e^{{(\delta-\alpha) (U_{0}+fL_{1})\over T_{c}}}]
+{L_{2}\over (U_{0}-fL_{2})}e^{{(\alpha -1)(U_{0}+fL_{1})\over T_{c}}-{(U_{0}+fL_{1})\delta \over T_{h}}-
{(\alpha -\delta) (U_{0}+fL_{1})\over T_{c}}}[e^{{(U_{0}-fL_{2})\over T_{c}}}-1]\\
G_{2}^{L}& = &{\gamma T_{c}L_{1}\over (U_{0}+fL_{1})}[e^{{(1-\alpha ) (U_{0}+fL_{1})\over T_{c}}}-1]
+{\gamma T_{c}L_{1}\over  (U_{0}+fL_{1})}e^{{(1-\alpha ) (U_{0}+fL_{1})\over T_{c}}
 +{ (U_{0}+fL_{1})\delta \over T_{h}}}
[e^{{(\alpha-\delta)  (U_{0}+fL_{1})\over T_{c}}}-1]
+{\gamma T_{h}L_{1}\over  (U_{0}+fL_{1})}e^{{(1-\alpha ) (U_{0}+fL_{1})\over T_{c}}}\nonumber \\ & &[e^{{\delta  (uo+fL_{1})\over
 T_{h}}}-1]+{\gamma T_{c}L_{2}\over (U_{0}-fL_{2})}
 e^{{(1-\alpha ) (U_{0}+fL_{1})\over T_{c}}+{ (U_{0}+fL_{1})\delta \over T_{h}}+ {(\alpha-\delta)  (uo+fL_{1})\over T_{c}}}
 [1-e^{{-U_{0}+fL_{2}\over T_{c}}}].\\ 
  H^{L}& = &T_{1}+T_{2}+T_{3}+T_{4}+T_{5}+T_{6}+T_{7}+T_{8}+T_{9}+T_{10} \\
  T_{1}& = &{L_{1}\gamma \over (U_{0}+fL_{1}) }[(1-\alpha )L_{1}+{T_{c}L_{1}\over  (U_{0}+fL_{1})}[
 e^{{ (U_{0}+fL_{1})(\alpha -1)\over T_{c}}}-1]],\\ 
 T_{2}& = &{L_{1}^{2}T_{c}\gamma \over (U_{0}+fL_{1}) ^{2}}e^{{ (U_{0}+fL_{1})(\alpha -1)\over T_{c}}}
      [e^{{(1-\alpha ) (U_{0}+fL_{1})\over T_{c}}}-1][1-e^{{- (U_{0}+fL_{1})\delta \over T_{h}}}],\\ 
 T_{3}& = &{L_{1}\gamma \over (U_{0}+fL_{1})}[\delta L_{1}+{T_{h}L_{1}\over (U_{0}+fL_{1})}[e^{{-(U_{0}+fL_{1})\delta \over T_{h}}}
-1]],\\
 T_{4}& = &{L_{1}^{2}T_{c}\gamma \over (U_{0}+fL_{1})^{2}}[1-e^{{(U_{0}+fL_{1})(1-\alpha )\over T_{c}}}]
 e^{{(U_{0}+fL_{1})(\alpha -1)\over T_{c}}-{\delta (U_{0}+fL_{1})\over T_{h}}}
 [e^{{-(U_{0}+fL_{1})(\alpha -\delta )\over T_{c}}}-1],\\ 
T_{5}& = &{L_{1}^{2}T_{h}\gamma \over (U_{0}+fL_{1})^{2}}[e^{{-(U_{0}+fL_{1})(\alpha -\delta )\over T_{c}}}-1]
[1-e^{{(U_{0}+fL_{1})\delta \over T_{h}}}]e^{{-(U_{0}+fL_{1})\delta \over T_{h}}},\\ 
T_{6}& = &{L_{1}\gamma \over (U_{0}+fL_{1})}[(\alpha -\delta )L_{1}+{T_{c}L_{1}\over (U_{0}+fL_{1})}
[e^{{-(U_{0}+fL_{1})(\alpha -\delta )\over T_{c}}}-1]],\\
 T_{7}& = &{L_{1}L_{2}T_{c}\gamma \over (U_{0}+fL_{1})(U_{0}-fL_{2})}[e^{{(U_{0}+fL_{1})(1-\alpha )\over T_{c}}}-1]
e^{{-(U_{0}+fL_{1})\over T_{c}}+{(U_{0}+fL_{1})\delta \over T_{c}}-{(U_{0}+fL_{1})\delta \over T_{h}}}
 [e^{{(U_{0}-fL_{2})\over T_{c}}}-1],\\
 T_{8}& = &{L_{1}L_{2}T_{h}\gamma \over (U_{0}-fL_{2}) (U_{0}+fL_{1})}[e^{{ (U_{0}+fL_{1})\delta \over T_{h}}}-1]
   [e^{{ (U_{0}-fL_{2})\over T_{c}}}-1]e^{{- (U_{0}+fL_{1})(\alpha -\delta )\over T_{c}}-{ (U_{0}+fL_{1})\delta \over T_{h}}
   },\\  
 T_{9}& = &{L_{1}L_{2}T_{c}\gamma \over (U_{0}+fL_{1})(U_{0}-fL_{2})}e^{{-(U_{0}+fL_{1})(\alpha -\delta )\over T_{c}}}
  [e^{{(U_{0}+fL_{1})(\alpha -\delta )\over T_{c}}}-1][e^{{(U_{0}-fL_{2})\over T_{c}}}-1],\\ 
   T_{10}& = &{-L_{2}\gamma \over(U_{0}-fL_{2})}[L_{2}-{L_{2}T_{c}\over (U_{0}-fL_{2})}[ e^{{(U_{0}-fL_{2})\over T_{c}}}-1]].\end{eqnarray} 
\end{widetext}
\begin{acknowledgments}
We would like to thank The Intentional Program in Physical Sciences, Uppsala University, Uppsala, Sweden for the support
they have been providing for our research group. MB would also like to thank the Kavli Institute for Theoretical Physics, 
University of california at Santa Barbara, for providing a conducive environment to write the major part of the paper during his visit there.
\end{acknowledgments}

\end{document}